\documentclass[aps,pra,showpacs,twocolumn]{revtex4}

\pdfoutput=1

\usepackage{graphicx}
\usepackage{dcolumn}
\usepackage{bm}
\usepackage{layout}
\usepackage{verbatim}

\usepackage{amsmath}
\usepackage{amssymb}
\newcommand{\beq}    {\begin{equation}}
\newcommand{\enq}    {\end{equation}}

\newcommand{\ket}[1]{\left|#1\right>}
\newcommand{\bra}[1]{\left< #1 \right|}

\begin{document}

\title{Spontaneous interlayer superfluidity in bilayer systems of cold polar molecules}

\author{Roman M. Lutchyn$^{1,2}$\footnote{Present address: Microsoft Research, Station Q, Elings Hall, University of California, Santa Barbara, CA 93106, USA}, Enrico Rossi$^{2}$\footnote{Present address: Department of Physics,
College of William and Mary, Williamsburg, VA 23187, USA}, S. Das Sarma$^{1,2}$}

\affiliation{$^{1}$ Joint Quantum Institute, Department of Physics,
             University of Maryland, College Park, Maryland 20742, USA\\
$^{2}$Condensed Matter Theory Center, Department of Physics,
             University of Maryland, College Park, Maryland 20742, USA}

\date{\today}

%--------------------------------

\begin{abstract}
%
%Quantum degenerate cold-atom gases provide a remarkable opportunity to
%study strongly interacting systems.
Recent experimental progress in
producing ultracold polar molecules with a net electric dipole
moment
%\cite{science_polar_KRB'08, PRL_polar_LiCs'08, PRL_polar_Rb2'08, Ospelkaus_natphys'08}
opens up new possibilities
to realize novel quantum phases governed by the long-range and
anisotropic dipole-dipole interactions. In this work we predict the
existence of experimentally observable novel broken-symmetry states
with spontaneous interlayer coherence in cold polar molecule bilayers. These
exotic states, which are manifestations of collective bilayer quantum entanglement,
appear due to strong repulsive interlayer interactions
and exhibit properties of superfluids, ferromagnets and excitonic
condensates.
\end{abstract}


\maketitle

%One of the outstanding features of the physics of cold atomic and
%molecular gases is the ability to simulate condensed matter systems
%with unprecedented control.
During the last decade we have observed
the spectacular progress in the realization of various quantum phases
using cold atoms. This progress has deepened our understanding of
various phenomena such as BCS-BEC crossover of
fermions
%\cite{Greiner_nature'03, Chin_science'04, Zwierlein_nature'05}
\cite{Greiner_nature'03}
and superfluid-Mott insulator phase transition
of bosons in an optical lattice
\cite{Greiner_nature'02}.
However, the
variety of quantum phases that can be realized in cold atom systems is
limited by the short-range nature of the interparticle
interactions. Recent progress in producing and manipulating
heteronuclear polar molecules
%\cite{science_polar_KRB'08, PRL_polar_LiCs'08, PRL_polar_Rb2'08, Ospelkaus_natphys'08}
\cite{science_polar_KRB'08}
provides an opportunity to
%take things
%to the next level and
realize a plethora of novel quantum phases of matter
governed by long-range interactions
%\cite{Micheli_natphys'06, wang_prl'06, wang_prl'07, buchler_prl'07, Chang_pra'09, cooper_prl'09}.
\cite{Micheli_natphys'06}.
This interesting prospect is made possible by the
fact that polar molecules have large electric dipole moments
associated with their rotational excitations, which lead to strong,
long-range and anisotropic dipole-dipole interactions. The
interactions between such polar molecules can be tuned using dc and ac
electric fields~\cite{Micheli_prb'07}. Below, we
concentrate on one intriguing aspect of fermionic polar molecule
systems - the possibility to realize bilayer superfluidity (or
equivalently bilayer $XY$ ferromagnetism) with spontaneous interlayer
coherence using interlayer molecular repulsion.

%Because of the fermionic nature of the molecules the many-body
%wavefuncion has to be antisymmetrized in order to satisfy the Pauli
%exclusion principle. As a consequence the interaction energy,
Because of the fermionic nature of the molecules even for
the spinless fermion case considered here, the interaction has an exchange component
in the layer (or pseudospin) index which drives the instability
towards the bilayer superfluid phase. The dominant contribution to the
exchange energy comes from the short distances (large momenta) where
the interlayer dipolar interaction is repulsive. Thus, the novel
collective bilayer state we predict arises from a repulsive
interaction in sharp contrast to all other superfluid quantum phases
discussed in cold atomic fermions where inter-particle attraction
leads to superfluidity.
%In this case, our predicted physics is close in spirit to that of interacting solid state systems where the bare
%interparticle interaction between electrons is {\it always} repulsive
%in nature.  In one respect, however, the phenomenon we discuss is
%qualitatively different from solid state phenomena where the
%interparticle interaction is coulombic (falling off as $1/r$) in
%contrast to the polar molecules with $1/r^3$ interaction.
%The dipolar
%$1/r^3$ interaction between the particles makes polar molecular gases
%a distinct new class of ``materials'' not accessible in solid state
%systems.
%The predicted interlayer superfluid state, where a $U(1)$
%symmetry is broken spontaneously, is rather strange because in the
In the symmetry broken interlayer coherent phase, the particle number in each
layer becomes indeterminate in spite of the interlayer single-particle
tunneling amplitude being almost zero.
%It is, as if, the bilayer
%system spontaneously develops a finite interlayer tunneling in spite
%of vanishing single particle tunneling!
%
Such a state is very analogous to an excitonic superfluid in which
excitons formed a quasiparticle in one layer ``binding'' to a quasihole in the other layer
condense into a phase coherent state.
Up to this date the clearest evidence for the realization of this
type of exciton superfluid state has been observed in semiconductor bilayers
in the Quantum Hall (QH) regime, in which the layers are immersed
in very high magnetic fields~\cite{Eisenstein_Nature'04}.
The unavoidable presence of disorder in solid state systems
as well as the nature of the measurement involving finite interlayer
tunneling~\cite{speilman_prl'00} cause complications in the pristine
realization of the interlayer superfluid phase in QH bilayers. As a result, vortices
and the Berezinskii-Kosterlitz-Thouless (BKT) transition, unambiguous
signatures of the interlayer coherent state, have not been observed yet in
solid state QH systems. The predicted polar molecule bilayer
superfluid phase should be more striking because of the lack of
disorder in cold atom systems, the tunability of the interaction
strength and the availability of experimental techniques allowing
imaging of vortices~\cite{Hadzibabic_nature'06}.
Moreover, by adding an optical lattice
potential in the $xy$-plane it is possible to modify the
single-particle dispersion of the particles and model various condensed matter systems.
For example, the bilayer cold-polar-molecule system with
honeycomb lattice potential will mimic the bilayer graphene
system. Thus, the realization of this interesting phase in cold atom
systems is of great importance for understanding the instabilities
driving the bilayer superfluidity and in general the physics of  exciton condensation.

\begin{figure}
%\begin{widetext}
 \begin{center}
\includegraphics[width=0.99\linewidth]{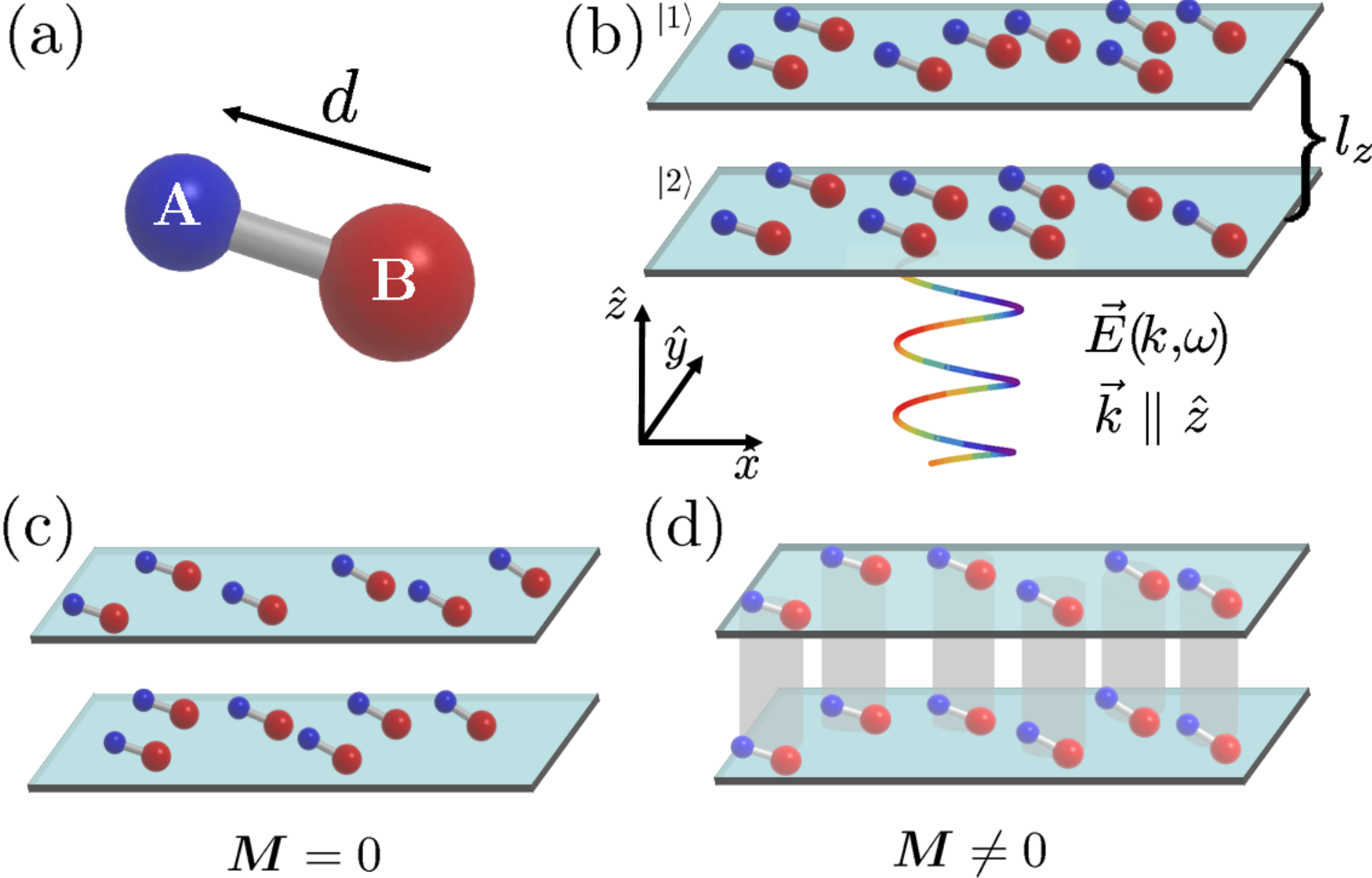}
  \caption{(Color online)  (a) Heteronuclear polar molecules AB with AB being
    $^{40}{\rm K}^{87}{\rm Rb}$, $^{7}{\rm Li}^{40}{\rm K}$, $^{6}{\rm
      Li}^{133}{\rm Cs}$. (b) Bilayer system of cold polar molecules
    in the presence of a circularly polarized ac electromagnetic field
    propagating along the $z$-direction. Schematic picture of
    different phases in the bilayers system: normal (c) and pseudospin
    ferromagnetic (d).}
 \label{fig:1}
 \end{center}
%\end{widetext}
\end{figure}

{\it Theoretical model. } Our starting point is the Hamiltonian for fermionic polar molecules
tightly confined along the $z$-direction by the laser field as shown
in Fig.~\ref{fig:1}b. We consider two clouds of polar molecules separated
by a distance $l_z$ much larger than the confinement length
$w_z$ of the molecules within each layer. When the confinement length
$w_z$ is much larger than the size of the polar molecules, the
rotational motion of the molecules is three-dimensional and is
described by a 3D rigid rotor Hamiltonian. Polar molecules have
permanent electric dipole moment $d$, which couples to internal
rotational degrees of freedom.  The dipole moment leads to long-range
interlayer and intralayer dipole-dipole interactions. The Hamiltonian
of the polar molecules $H$ reads ($\hbar=k_B=1$)~\cite{Micheli_prb'07}
\begin{align}
\!H&\!=\!\sum_i\! \left(\frac{\bm p_i^2}{2 m}\!+\!B \bm
J_i^2\!\right)\!+\! \sum_{ij}\!\frac{\bm d_i \bm d_j \!-\!3 (\bm d_i
  \!\cdot\! \hat {\bm r}_{ij})(\bm d_j\!\cdot\! \hat {\bm  r}_{ij})}{2r_{ij}^3},
\end{align}
where $\bm p = (p_x,p_y)$ is the center-of-mass momentum of a molecule
with mass $m$, $r_{ij}$ is the distance between two molecules, $B$ is
the effective rotational energy and $\bm J=(J_x, J_y, J_z)$ is the
angular momentum operator. The rotational eigenstates are $\ket{J,
  M_J}$ with $J$ and $M_J$ denoting the total internal angular
momentum and its projection on the quantization axis, respectively.

%%%%%%%%%%%%%%%%%%
%\paragraph{Many-body physics. }
\begin{figure}
%\begin{widetext}
 \begin{center}
\includegraphics[width=3.4in, angle=0]{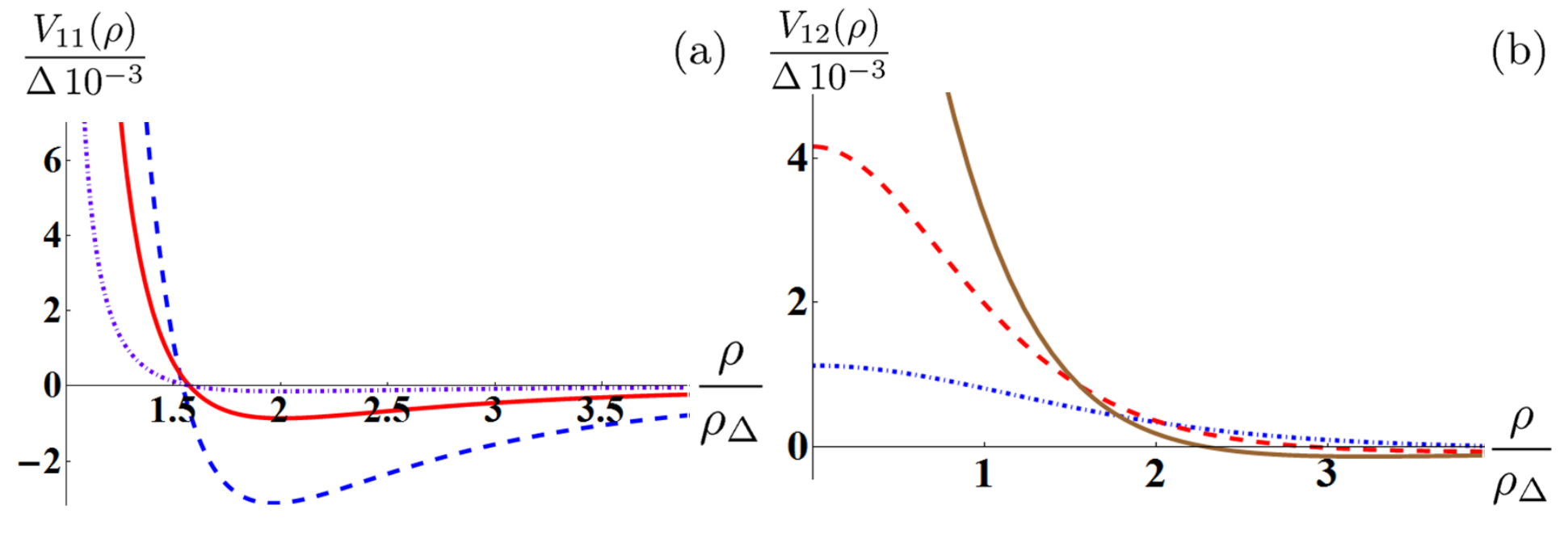}
  \caption{(Color online) (a) Intralayer Born-Oppenheimer potential for polar
    molecules. The dashed(blue), solid(red) and dot-dashed(violet)
    lines correspond to %different Rabi frequencies:
    $\Omega_R/\Delta=1/4$, $\Omega_R/\Delta=1/8$ and
    $\Omega_R/\Delta=1/20$, respectively. %Inset: two-body energy
    %levels as a function of the distance.
    %Avoided level crossing at
    %$\rho\sim \rho_{\Delta}$ leads to short-range repulsion between
    %the molecules allowing to stabilize the system.
    (b) Interlayer
    Born-Oppenheimer potential for $\Omega_R/\Delta=1/8$. The
    dash-dot(blue), solid(red) and dashed(brown) lines correspond to
    $l_z/\rho_{\Delta}=3, 2,1.5$, %, $l_z/\rho_{\Delta}=2$ and
    $l_z/\rho_{\Delta}=1.5$, respectively. For typical interparticle
    distances considered here the interlayer interaction is
    repulsive. %Inset shows the agreement between perturbative (dashed
    %red line) and exact (solid green line) Born-Oppenheimer
    %potentials.
    Here $\Omega_R/\Delta=1/8$ and $l_z/\rho_{\Delta}=3$.}
 \label{fig:2}
 \end{center}
%\end{widetext}
\end{figure}
%%%%%%%%%%%%%%%%%%%

The orientation of the dipole moments can be controlled with dc and ac
electric fields Fig.~\ref{fig:1}b, \cite{Micheli_prb'07}.  The
transition dipole moment between the states with $J=0$ and $J=1$ is
$d_t \equiv |\bra{0, 0}\bm d\ket{1, M_J}|=d/\sqrt{3}$ with $M_J=0,\pm
1$.  A circularly polarized ac electric field $\bm E_{\rm ac}(t)$
propagating along $z$-direction, see Fig.~\ref{fig:1}b, drives
transitions between the rotational states $\ket{0,0}$ and $\ket{1,1}$
with Rabi frequency $\Omega_R=d_t E_{\rm ac}$. If the frequency of the
field $\omega$ is close to the transition frequency $\omega_0=2B$
between the states $\ket{0,0}$ and $\ket{1,1}$ ({\it i.e.} the
detuning $\Delta=\omega-\omega_0\ll \omega_0$), the leading effect of
the electric field is to mix these two states. Within the rotating
wave approximation, the dressed states are given by
$\ket{\pm}=\alpha_{\pm}\ket{0,0}\pm \alpha_{\mp}e^{-i\omega
  t}\ket{1,1}$, where $\alpha_+=-\Gamma/\sqrt{\Gamma^2+\Omega_R^2}$,
$\alpha_-=\Omega_R/\sqrt{\Gamma^2+\Omega_R^2}$ and
$2\Gamma=\Delta+\sqrt{\Delta^2+4\Omega_R^2}$~\cite{cooper_prl'09, Micheli_prb'07}.
Polar molecules can be prepared
in the internal state $\ket{+}_i$ by an adiabatic switching of the
microwave field. In this case, the effective interaction between polar
molecules $V_{\rm eff}(r)$ is given by the dressed Born-Oppenheimer
potential adiabatically connected to the state $\ket{+}_i\otimes
\ket{+}_j$, see Fig.1 of the supplementary material. %(solid blue line in the inset of Fig.~\ref{fig:2add} in supplementary material).
At large distances the dipolar interaction can be obtained perturbatively
by first calculating the effective dipole moment $\bra{+}\bm d
\ket{+}=d_{\rm eff }(\cos \omega t, \sin \omega t,0)$ with $d_{\rm
  eff}=-\sqrt{2} \alpha_+ \alpha_- d_t$. The time-averaged interaction
between dipoles in layers $\lambda$ and $\lambda'$ takes the form
\begin{align}\label{eq:V_eff}
V^{\lambda \lambda'}_{\rm eff}(\rho)= d_{\rm eff}^2 \left(
\frac{1}{\left(z_{\lambda \lambda'}^2+\rho^2\right)^{\frac{3}{2}}}
-\frac 3 2 \frac{\rho^2}{\left(z_{\lambda
    \lambda'}^2+\rho^2\right)^{\frac{5}{2}}}\right),
\end{align}
where $\bm \rho=(x,y)$ is the 2D coordinate, and $z_{\lambda
  \lambda'}=l_z$ for $\lambda \neq \lambda'$ and zero otherwise. At
short distances, when the dipolar interaction energy is comparable
with the detuning, the above perturbative treatment breaks down. In
order to find the Born-Oppenheimer potential at short distances $\rho
\leq \rho_{\Delta}\equiv (d_t^2/\Delta)^{\frac 1 3}$, it is necessary
to account for all couplings between different angular momentum
channels within the $J=0,1$ manifold
%\cite{Micheli_prb'07, gorshkov_prl'08}.
\cite{Micheli_prb'07}.
The exact Born-Oppenheimer potentials are shown in
Fig.~\ref{fig:2}.  One can notice that the effective intralayer
dipole-dipole interaction between polar molecules prepared in the
state $\ket{+}$ becomes repulsive at $\rho \sim \rho_{\Delta}$ due to
the presence of avoided crossings with other field-dressed levels~\cite{cooper_prl'09}. For
$^{6}{\rm Li}^{133}{\rm Cs}$ molecules, typical parameters are $d
\approx 6.3\rm D$, $B\approx 6$GHz, and $\Delta \approx 10$MHz
yielding the length scale $\rho_{\Delta}\approx 50$nm that is much
larger than the characteristic scale of dipole-dipole interactions
$\rho_B\!=\!(d^2/B)^{1/3}\!\sim\! 1$nm, which sets the short-range
cutoff. Thus, for typical densities of polar molecules considered here,
$n_0\! \sim \! 10^7$cm$^{-2}$, the ac electric field shields the
molecules from short-range inelastic collisions and prevents from the
collapse of the system
%\cite{Napolitano_pra'97, cooper_prl'09, Micheli_prb'07, gorshkov_prl'08},
\cite{Napolitano_pra'97, cooper_prl'09, Micheli_prb'07},
see supplementary material.

Henceforth, we consider the dilute gases of polar molecules, where the
interparticle distance is larger than $\rho_{\Delta}$, \emph{i.e.}
$l_z, n_0^{-1/2} \gg \rho_{\Delta}$, and the interaction between
particles is given by the dressed Born-Oppenheimer potentials shown in
Fig.~\ref{fig:2}. In order to avoid unwanted inelastic collisions
leading to the decay of the molecules in an s-wave channel, we assume
the molecules to be spin-polarized. In this limit, the effective second-quantized Hamiltonian of the
bilayer system takes the form
\begin{align}\label{eq:manybodyHamiltonian}
\!H\!&\!=\!\sum_{k \lambda} ( \varepsilon(k)\!-\!\mu_\lambda)
c^\dag_{k\lambda} c_{k \lambda}\\ &\!+\!\frac 1 2\!
\sum_{q,k,k',\lambda \lambda'}\!V_{\rm eff}^{\lambda\lambda'}(q)
c^\dag_{k+q \lambda} c^\dag_{k'-q \lambda'} c_{k' \lambda'} c_{k
  \lambda},\nonumber
\end{align}
where $c_{k\lambda}$ and $c^\dagger_{k\lambda}$ are the fermion
creation and annihilation operators for a molecule with momentum $\bm
k$ in layer $\lambda$. The strength of the dipolar interactions can be
characterized by the dimensionless parameter
%\begin{align}
$r_s=d_{\rm eff}^2 m \sqrt{n_0}/2\pi$.
%\end{align}
As $r_s$ is increased, the bilayer system becomes susceptible to
various instabilities driven by the dipolar interactions. Here we
concentrate on the instabilities induced by the interlayer
interactions and neglect the instabilities induced by the intralayer
interactions, see supplementary material for the justification of such
approximations.

To understand the interlayer instability, it is convenient to draw an
analogy with ferromagnetism and introduce pseudospin-1/2 operators
$2\bm {\hat m}_i= c^\dag_{i\lambda}\bm
\sigma_{\lambda\lambda'}c_{i\lambda'}$. The spinors $\ket{\uparrow}$
and $\ket{\downarrow}$ represent the states in which molecules are in
layer 1 or 2, respectively. When $l_z \gg k_F^{-1}$, with $k_F=\sqrt{4\pi n_0}$ being the Fermi momentum, the
molecules in different layers are uncorrelated, and the many-body
state of the system is given by
\begin{align}
\ket{\Psi_{\rm N}}=\prod_{k \leq k_F} c_{k1}^\dag \prod_{k' \leq k_F}
c_{k'2}^\dag\ket{0}.
\end{align}
For equal densities in the layers $n_1=n_2=n_0$, the total
magnetization $\bm M$ is zero, $\bm M=\bra{\Psi_{\rm N}}\sum_i \bm
{\hat m}_i\ket{\Psi_{\rm N}}=0$, similar to paramagnets. The normal
state $\ket{\Psi_{\rm N}}$ minimizes the kinetic energy at the expense of
the potential energy, which is at its maximum. When the interlayer
distance becomes smaller than $k_F^{-1}$, the potential energy becomes
large and at some point starts to dominate over the kinetic energy. In
this case, the system favors the state in which fermions in different
layers are correlated in a way that minimizes the interaction energy,
\emph{i.e.} the molecule in layer 1 is coupled to a ``hole" in layer
2.  At the mean-field level such correlations are captured by an order
parameter $\Delta_{12}\propto \langle c^\dag_{k1} c_{k2} \rangle \neq
0$.  Since the product wavefunction $\Psi_{\rm N}$ does not have such
entanglement between the layers, the bilayer system should undergo a
quantum phase transition as a function of the distance $l_z$ or the
strength of the dipole moment $d_{\rm eff}$. The many-body
wavefunction minimizing the interaction energy takes the form
\begin{align}\label{eq:pseudospin_wave}
\ket{\Psi_{\rm FM}}=\prod_{k \leq \sqrt{2}k_F} \left(
\frac{c^\dag_{k1}+e^{i\varphi} c^\dag_{k2}}{\sqrt{2}}\right) \ket{0}.
\end{align}
In this entangled state
%reveals the true nature of quantum mechanics
%since now
the state of the molecule is given by the coherent
superposition of the amplitudes in different layers. Thus, even in the
absence of tunneling, the molecule layer index becomes
uncertain. Using the spin analogy, the state $\ket{\Psi_{\rm FM}}$ has
non-zero magnetization $\bm M=\!\!\bra{\Psi_{\rm FM}}\! \sum_i \bm
{\hat m}_i\!\ket{\Psi_{\rm FM}}\!\neq\!0\!$ with $\bm M$ lying in the
$xy$-plane. Similar to superconductors, this ferromagnetic state
$\ket{\Psi_{\rm FM}}$ spontaneously breaks $U(1)$ symmetry and
develops interlayer coherence. In this state the phase difference
between different layers $\varphi$ is well defined and the number of
molecules in each layer fluctuates satisfying the uncertainty
relations $\Delta m_z \Delta \varphi \geq 1/2$~\cite{moon_prb'95}
despite the absence of interlayer tunneling in
Eq.~\eqref{eq:manybodyHamiltonian}.

The phase diagram between the two competing states - normal
$\ket{\Psi_{\rm N}}$ and pseudospin ferromagnetic $\ket{\Psi_{\rm
    FM}}$ - can be obtained using variational mean-field
calculation~\cite{zheng_prl'97}. The total energy per area $\mathcal
A$ of the bilayer system in the normal phase is
\begin{align}
\frac{E_{\rm N}}{\mathcal{A}}=\frac{2\pi n_0^2}{m}\left(1+r_s
\frac{Z(2k_F\rho_{\Delta})}{k_F \rho_\Delta \frac{d_{\rm
      eff}^2}{d_t^2}} \! \right),
\end{align}
where the dimensionless function $Z(a)$, which describes intralayer
interaction, is defined as
\begin{align}\label{eq:Z}
Z(a)\!=\! \frac{ 32}{\sqrt \pi} \int_0^1 \! x dx f(x) \! \frac{
  \left[V_{\rm eff}^{(11)} (0)\!-\!V_{\rm eff}^{(11)} (a x) \!\right]
}{\Delta \rho_{\Delta}^2}
\end{align}
with $f(x)=\arccos(x)-x\sqrt{1-x^2}$. The momentum
dependence of the intralayer interaction potential $V_{\rm
  eff}^{(11)}(q)$ is shown in Fig. 1 (b) of the supplementary material. One can notice that the contribution of the interlayer
Hartree term is zero here because $V^{12}_{\rm eff}(q)=d_{\rm eff}^2
\pi qe^{-ql_z}$ goes to zero as $q \rightarrow 0$. To calculate the
energy of the system in the interlayer coherent state $\ket{\Psi_{\rm
    FM}}$ we first introduce the order parameter
\begin{align}
\Delta_{12}(k)= \frac 1 2 \sum_{q} V_{12}(q) e^{-i \varphi}
\left\langle \! c^\dag_{1}\left(k\!+\!q\right)\! c_{2}\left(k+q\right)
\! \right \rangle,
\end{align}
which takes into account interlayer correlations.  The order parameter
is obtained by numerically solving the self-consistent equation above,
subject to the total particle number conservation constraint.  Because
of the dipolar nature of the interaction $\Delta_{12}(k)$ has momentum
dispersion. The dependence of $\Delta_{12}(0)$ on $d_{\rm eff}$
and $l_z$ obtained self-consistently is shown in
Fig.~\ref{fig:delta}~(a).
At the mean-field level, the
Hamiltonian~\eqref{eq:manybodyHamiltonian} can be diagonalized using a
Bogoliubov transformation yielding the many-body variational
wavefunction \eqref{eq:pseudospin_wave}.  For sufficiently large
interactions the lowest energy state of the system corresponds to the
pseudospin ferromagnetic state fully polarized in the
$xy$-plane~\cite{zheng_prl'97}, see also supplementary material. The
total energy of the system per area in the interlayer coherent state
is given by
\begin{align}
\frac{E_{\rm FM}}{\mathcal A}=\mathcal{E}_0 \left[1-r_s
  F\!\left(\!\sqrt{8} k_{\rm F}l_z\! \right) +
  \frac{r_s}{2}\frac{Z(\sqrt 8 k_F \rho_{\Delta})}{k_F \rho_{\Delta}
    \frac{d_{\rm eff}^2}{d_t^2}}\right],
\end{align}
where $\mathcal{E}_0=4\pi n_0^2/m$ and the function $F(a)$ describing
the interlayer exchange contribution reads
\begin{align}\label{eq:f}
\!\!\!F(a)\!=\! 32 \sqrt{2\pi} \!\! \int_0^1\! x^2 \! dx \!
\left(\arccos(x)\!-\!x\sqrt{1\!-\!x^2}\right)\!e^{-a x}.
\end{align}
The energy difference between normal and ferromagnetic phase $\Delta
E/{\mathcal A} = (E_{\rm FM} - E_{\rm N})/{\mathcal A}$ determines the
mean-field phase diagram for the bilayers of polar molecules shown in
Fig.~\ref{fig:delta}.
\begin{figure}
 \begin{center}
 \includegraphics[width=0.98\linewidth]{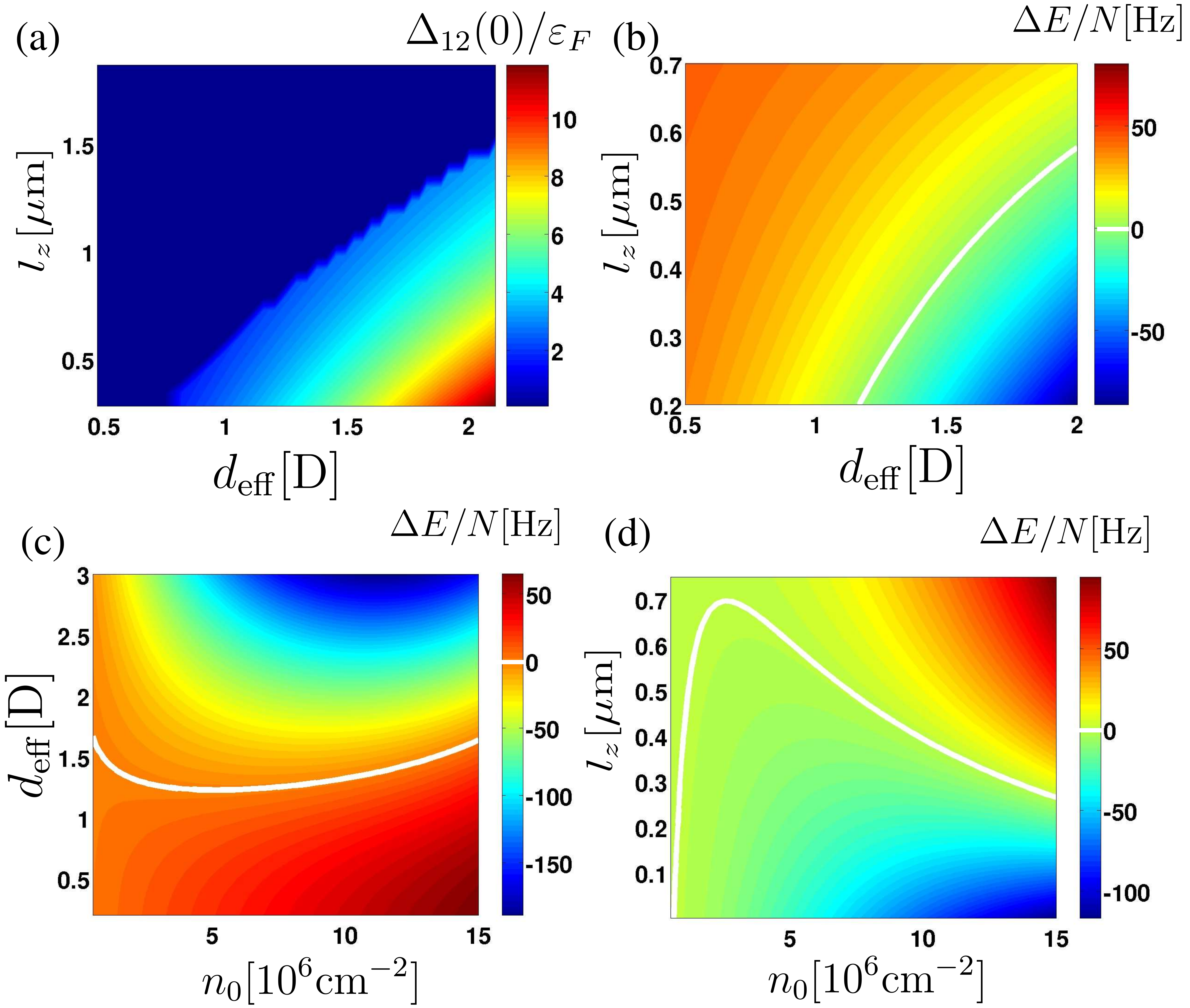}
  \caption{(Color online)
    %The dependence of the order parameter $\Delta_{12}$
    %and energy difference $\Delta E$ on the physical parameters of the
    %system.
    The dependence of $\Delta_{12}(0)$ and $\Delta E$ on
    $l_z$ and $d_{\rm eff}$ is shown in (a) and (b),
    respectively. Here we used $\rho_{\Delta}=50$nm,
    $n_0=10^7$cm$^{-2}$ and $m=139m_p$ with $m_p$ being the proton
    mass.  The dependence of $\Delta E$ on $n_0$ and $d_{\rm eff}$ at
    fixed $l_z=0.3\mu{\rm m}$ and on $n_0$ and $l_z$ at fixed $d_{\rm
      eff}=1.5$~D is plotted in (c) and (d), respectively.  }
  \label{fig:delta}
 \end{center}
\end{figure}

The long wavelength Hamiltonian describing the phase
fluctuations is
\begin{align}\label{eq:Goldstone}
H=\frac 1 2 \int d^2 \bm r \rho_s |\nabla \varphi |^2,
\end{align}
where $\rho_s$ is the ``spin stiffness", which is the result of the
loss of interaction energy due to the spatial variations of
order parameter phase $\varphi$. The effective $XY$ model defined by
Eq.~\eqref{eq:Goldstone} undergoes BKT transition associated with
unbinding of vortex pairs at the temperature $T_{\rm BKT} \approx \pi
\rho_s/2$. The ``spin stiffness" can be calculated within linear
response theory, which yields the result $\rho_s=n_0/2m$, similar to
the one in superfluids.  Thus, the BKT transition in the bilayer
system of polar molecules occurs at the temperature
%\begin{align}
$T_{\rm BKT}\approx \varepsilon_F/8$ with $\varepsilon_F=k_F^2/2m$,
that for $n_0\sim 10^7$cm$^{-2}$ corresponds to a temperature that
should be accessible in the near future \cite{ye_private_2009}.
%\end{align}
In the pseudospin ferromagnetic phase vortices are correlated in
different layers, and the BKT transition in the bilayer system can be
detected by imaging vortices using matter wave heterodyning
techniques~\cite{Hadzibabic_nature'06}.

In summary, we predict an unusual broken-symmetry phase with
spontaneous interlayer coherence in a bilayer system of cold polar
molecules. Our main findings, summarized in the phase diagram shown in
Fig.~\ref{fig:delta}, indicate that the experimental observation of
such phase requires low densities of cold polar molecules $n_0\sim
10^7$cm$^{-2}$, realistic dipole moments $d_{\rm eff}\sim1$D, and
reasonably low temperatures $T\sim 1$nK. Given that for these
parameters the inelastic decay rate is small~\cite{cooper_prl'09}, the detection of this
exotic many-body state should be within the experimental reach in the
near future.
The novel phase we predict is an interlayer entangled state, arising
from the repulsive part of the dipolar interaction and exhibiting
superfluidity (or, equivalently, $XY$ pseudospin ferromagnetism) between
the layers rather than within the layers.

{\it Acknowledgements.} We thank P. S. Julienne, I. Spielman and T. Porto,
and most particularly D.-W. Wang, for helpful discussions.
This work is supported by US-AFOSR-MURI and NSF-JQI-PFC.

\vspace*{-0.2in}

\newpage

\title{SUPPLEMENTARY INFORMATION: \\
Spontaneous interlayer superfluidity in bilayer systems of cold polar molecules}

\author{Roman M. Lutchyn$^{1,2}$\footnote{Present address: Microsoft Research, Station Q, Elings Hall, University of California, Santa Barbara, CA 93106, USA}, Enrico Rossi$^{2}$\footnote{Present address: Department of Physics,
College of William and Mary, Williamsburg, VA 23187, USA}, S. Das Sarma$^{1,2}$}

\affiliation{$^{1}$ Joint Quantum Institute, Department of Physics,
             University of Maryland, College Park, Maryland 20742, USA\\
$^{2}$Condensed Matter Theory Center, Department of Physics,
             University of Maryland, College Park, Maryland 20742, USA}

\date{\today}

\maketitle

\section{SUPPLEMENTARY INFORMATION}
%=========================================================

In this document we elaborate on the analysis of the various instabilities in the bilayer system of cold polar molecules. We begin with the discussion of the role of the intralayer interactions in stabilizing the cold-polar-molecule Fermi gas.

\subsection{Stability analysis of the single layer}
\begin{figure}{!hb}
%\begin{widetext}
 \begin{center}
\includegraphics[height=3in, angle=90]{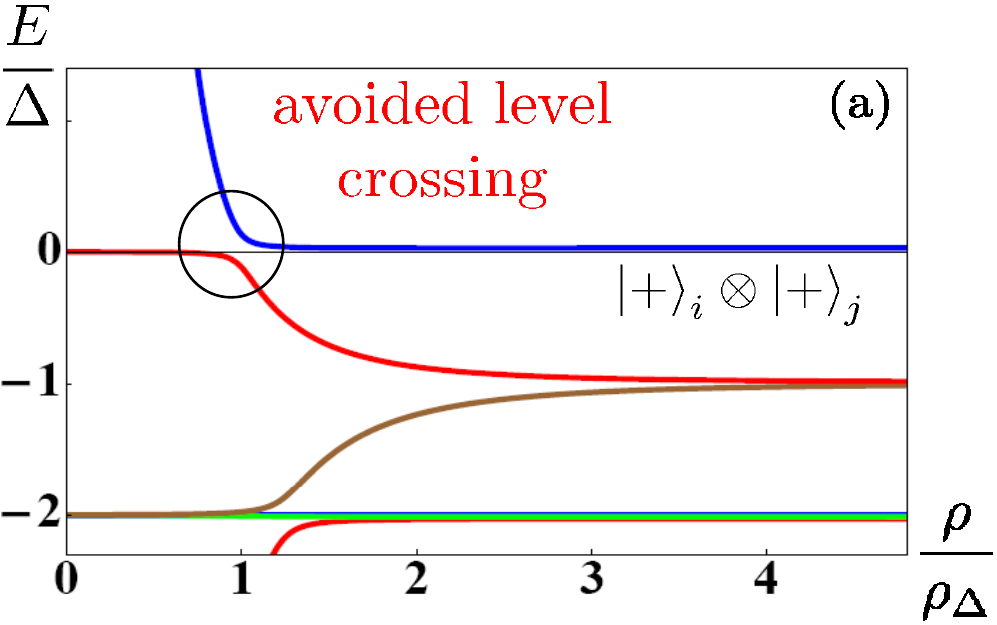}
  \caption{Two-body energy levels as a function of the distance for $\Omega_R/\Delta=1/8$.}
 \label{fig:BO}
 \end{center}
%\end{widetext}
\end{figure}

The Hamiltonian for a single layer 2D Fermi gas of cold polar molecules reads
\begin{align}\label{eq:singlelayerHamiltonian}
\!H\!&\!=\!\sum_{k} ( \varepsilon(k)\!-\!\mu) c^\dag_{k\lambda} c_{k}+\!\frac 1 2\! \sum_{q,k,k'}\!V_{\rm eff}^{(11)}(q) c^\dag_{k+q} c^\dag_{k'-q} c_{k'} c_{k}.
\end{align}
The intralayer interaction  is given by the dressed Born-Oppenheimer potential $V_{11}(\rho)$~\cite{Micheli_prb'07, gorshkov_prl'08}, which has a short-distance repulsive core and an attractive tail at large distances as shown in Fig.~\ref{fig:intra}a of the main text. The crossover between these two regimes occurs at the length scale $\rho_{\Delta}$, which is determined by the detuning $\Delta$, see Fig.~\ref{fig:BO}. Thus, by changing the detuning frequency one can control the strength of the intralayer interaction.

%We note that it is implicitly assumed here that van der Waals interactions are negligible. This approximation is justified because polar-molecule densities considered here $n_0\sim 10^7$cm$^{-2}$ are small and, thus, typical interparticle distances $\sim 1\mu$m are large. However,

The energy of the single layer in the Hartree-Fock approximation~\cite{chan_arxiv'09} is given by
\begin{align}
\frac{E^{(\rm sl)}_{\rm t}}{\mathcal{A}}&=\sum_k \frac{k^2}{2m}n_k \!+\! \frac 1 2 \sum_{k,q}\left[V_{\rm eff}^{(11)} (0)\!-\!V_{\rm eff}^{(11)} (q)\right] n_{k} n_{k\!+\!q} \nonumber\\
&=\frac{\pi n_0^2}{m} \!+\! \frac{ 1}{2}\! \int \! \frac{d^2\bm q}{(2\pi)^2} \! \left[V_{\rm eff}^{(11)} (0)\!-\!V_{\rm eff}^{(11)} (q)\right] I(q, k_F),
\end{align}
where $n_k=\Theta(k_F-k)$ with $\Theta(x)$ being the Heaviside step function. The
%we used $\frac{1}{V} \sum_{k} n_{k} \!=\! k_F^2/4\pi=n$ and
dimensionless function $I(q, x)$ is given by
\begin{align}
\!I(q, x)&\!\!=\!%\frac{1}{V} \sum_{k} n_{k}n_{k\!+\! q}\!\!=
\!\! \int \!\! \frac{d^2\bm k}{(2\pi)^2} \Theta(x\!-\!\sqrt{k^2\!+\!q^2\!+\!2\bm k \bm q})\Theta(x \!-\!k) \\
&\!=\!\frac{x^2}{2\pi^2}\Theta(2x\!-\!q) \! \left[\arccos\left(\! \frac{q}{2x}\!\right)\!-\!\left(\frac{q}{2x}\right)\!\sqrt{1\!-\!\left(\!\frac{q}{2x}\!\right)^2}\right], \nonumber
\end{align}
and the momentum dependence of the intralayer potential $V_{\rm eff}^{(\lambda \lambda)} (0)-V_{\rm eff}^{(\lambda \lambda)} (q)$ is shown in Fig.~\ref{fig:intra}b. Then,
%\end{align}
the total energy of the system becomes
\begin{align}\label{eq:normal_single}
\frac{E^{(\rm sl)}_{\rm t}}{\mathcal{A}}\!&=\!\frac{\pi n_0^2}{m} + \frac{ 8}{\pi}n_0^2\! \int_0^1 \! x dx \! \left[V_{\rm eff}^{(11)} (0)\!-\!V_{\rm eff}^{(11)} (2k_F \rho_{\Delta} x) \right] \! f(x)\nonumber\\
\!&\!=\! \frac{\pi n_0^2}{m}\! \left( \!1\! +\! r_s \frac{Z(2k_F\rho_{\Delta})}{k_F \rho_\Delta \frac{d_{\rm eff}^2}{d_t^2}} \! \right)
\end{align}
with $r_s=d_{\rm eff}^2 m \sqrt{n_0}/2\pi$.
%the momentum dependence of the intralayer interaction potential is shown in Fig.~\ref{fig:intra}b,
The dimensionless function $Z(a)$ is defined as
\begin{align}\label{eq:Z}
Z(a)\!=\! \frac{ 32}{\sqrt \pi} \int_0^1 \!\! f(x) x dx \! \frac{ \left[V_{\rm eff}^{(11)} (0)\!-\!V_{\rm eff}^{(11)} (a x) \!\right] }{\Delta \rho_{\Delta}^2},
\end{align}
where
%\begin{align}
$f(x)=\arccos(x)-x\sqrt{1-x^2}$. Since the integral~\eqref{eq:Z} involves both the contribution from the attractive tail and repulsive core of the potential $V_{\rm eff}^{(11)}(\rho)$ (see Fig~\ref{fig:intra}), the intralayer interaction energy depends on the parameter $k_F \rho_{\Delta}$. For typical densities considered here $n_0\! \approx\!  10^7$cm$^{-2}$ and $\rho_{\Delta}\!\approx\! 50$nm ({\it i.e.} $k_F\rho_{\Delta}\approx 0.06$), the compressibility of the system $\kappa^{-1}=\frac{n_0^2}{\mathcal{A}} \frac{\partial^2 E^{(\rm sl)}_{\rm t}}{\partial n_0^2}$ is positive, see Fig~\ref{fig:intra}c,  and thus, the ac electric field allows one to stabilize the Fermi gas and prevent phase separation.

\subsection{Intralayer superfluid instability.}

\begin{figure}
%\begin{widetext}
 \begin{center}
\includegraphics[width=1.0\linewidth]{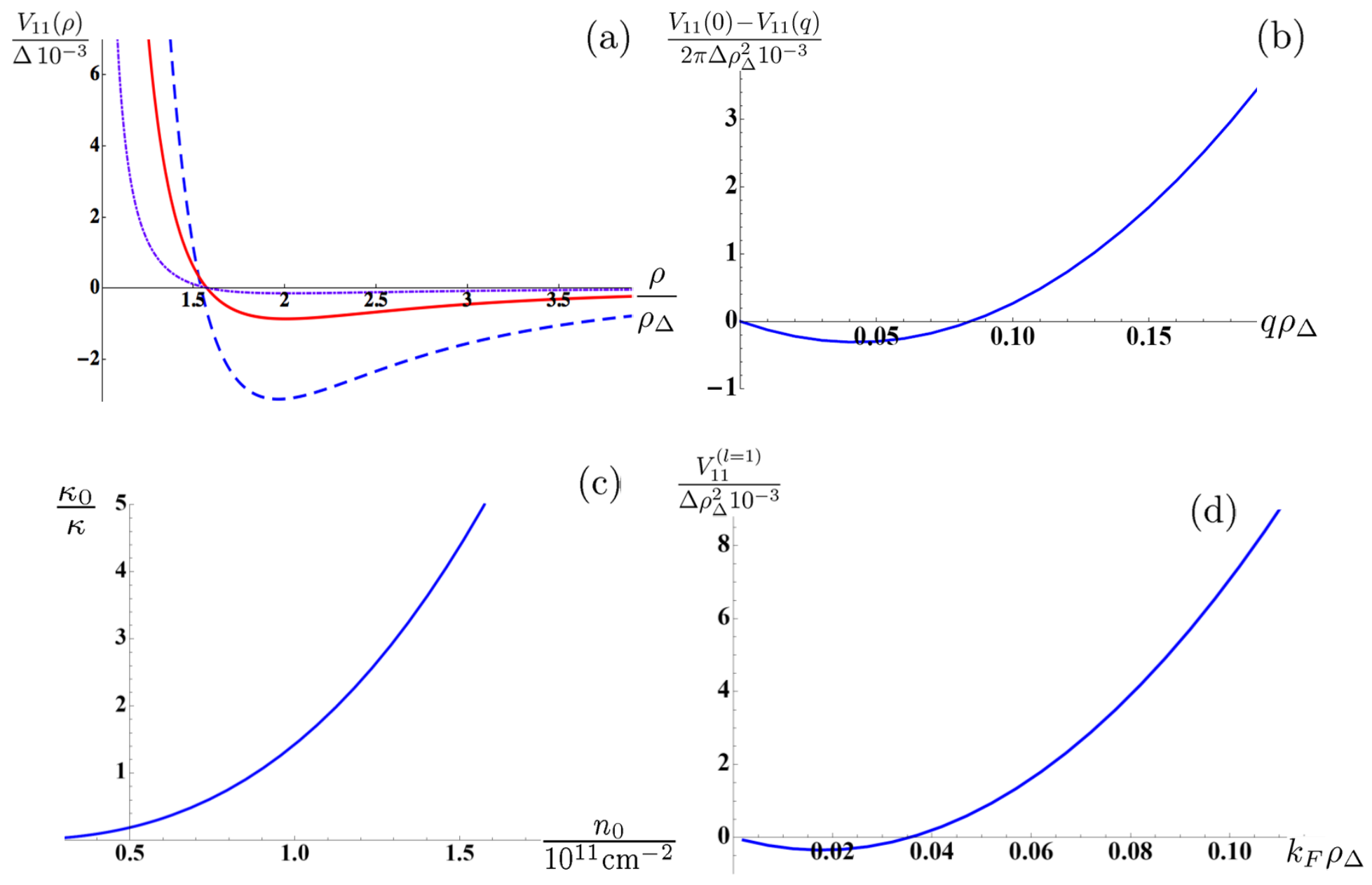}
  \caption{(a) Intralayer Born-Oppenheimer potential for polar molecules. The dashed(blue), solid(red) and dot-dashed(violet) lines correspond to different Rabi frequencies: $\Omega_R/\Delta=1/4$, $\Omega_R/\Delta=1/8$ and $\Omega_R/\Delta=1/20$, respectively. (b) Fourier transform of the intralayer Born-Oppenheimer potential for $\Omega_R/\Delta=1/8$. (c) The compressibility of the single-layer cold polar Fermi gas as a function of density $n_0$. Here $\kappa_0=m/\pi$. (d) The dependence of the p-wave potential $V^{(l=1)}$ on the dimensionless parameter $k_F\rho_{\Delta}$.}
 \label{fig:intra}
 \end{center}
%\end{widetext}
\end{figure}

We now discuss the possibility of the superfluid instability of the Fermi
surface in a single layer of cold polar molecules.
%Thus, the bilayer system of polar molecules provides a unique possibility to study the
%interplay of interlayer pseudospin ferromagnetic and intralayer superfluid instabilities.
It was shown in Ref.~\cite{cooper_prl'09}
that the intralayer interaction can lead to p-wave pairing. The weak coupling BCS theory developed in Ref.~\cite{cooper_prl'09} is valid for $r_s \ll 1$,
and the evaluation of the superfluid $T_c$ for typical parameters considered here ($r_s \lesssim 1$) requires developing a strong
coupling theory of superfluidity, which is beyond the scope of the present paper. However, we note that by tuning the
ratio $\Omega_R/\Delta$ the depth of the attractive potential can be decreased as shown in Fig.~\ref{fig:intra}a indicating
that it is possible to substantially suppress the intralayer superfluid $T_c$ or eliminate such instability altogether. To demonstrate this, we calculate the intralayer interaction
in the p-wave channel $V^{(l=1)}$
\begin{align}
V^{(l)}\equiv V^{(l)}(k_F,k_F)=2\pi \int_0^{\infty}\rho d\rho J^2_l(k_F \rho) V^{(11)}(\rho),
\end{align}
which can be controlled by changing $\Omega_R/\Delta$. Here $J_l(k_F \rho)$ is the Bessel function of order $l$. Due to the short range repulsion the potential $V^{(l=1)}$ changes sign as a function of $k_F \rho_{\Delta}$, see Fig.~\ref{fig:intra}d. For the relevant parameters considered here $\Omega_R/\Delta=1/8$ and $k_F \rho_{\Delta}\approx 0.06$, the potential $V^{(l=1)}$ is positive, and the analysis of the spontaneous bilayer coherence neglecting the intralayer instabilities in the particle-particle channel is justified. We note here that there is no inconsistency with the conclusion of Cooper and Shlyapnikov~\cite{cooper_prl'09} since for the parameters used in Ref.~\cite{cooper_prl'09} ($\Omega_R/\Delta=1/4$ and $\rho_{\Delta}=30$nm) the potential is $V^{(l=1)}$ is indeed negative leading to the p-wave superfluidity.

\subsection{Effect of intralayer interactions on the pseudo-spin instabilities}

In this section we provide details on the calculation of the energy of the bilayer system of cold polar molecules within variational mean-field theory. We begin by analyzing the total energy of the system per area in the normal phase described by the wavefunction $\ket{\Psi_{\rm N}}$ (see Eq.~(4) of the main text):
\begin{align}
\!&\!\frac{E_{\rm N}}{\mathcal{A}}\!=\!\frac{2\pi n_0^2}{m}\!+\!\frac{ 16}{\pi}n_0^2\! \int_0^1 \! x dx \! \left[V_{\rm eff}^{(11)} (0)\!-\!V_{\rm eff}^{(11)} (2k_F \rho_{\Delta} x) \right] \! f(x),\nonumber\\
\!&\!=\! \frac{2\pi n_0^2}{m}\! \left( \!1\! +\! r_s \frac{Z(2k_F\rho_{\Delta})}{k_F \rho_\Delta \frac{d_{\rm eff}^2}{d_t^2}} \! \right),
\end{align}
The energy $E_{\rm N}$ essentially follows from Eq.\eqref{eq:normal_single} multiplied by the number of layers.

We now discuss the energy of the bilayer system in the superfluid state. Assuming that the dominant contribution to the energy comes from interlayer interactions, we introduce the mean field $\Delta_{12}(k)$ (see Eq.~(8)  of the main text) and neglect for now the intralayer interactions. At the mean-field level, the Hamiltonian
\begin{align}
H_{\rm MF}\!=\!\sum_{k \lambda} \left(\frac{k^2}{2m}\!-\!\mu \! \right) \! c_{k \lambda}^\dag \! c_{k \lambda}\!-\!\sum_k \! \left[\Delta_{12}(k) c_2^\dag(k)c_1(k)\!+\!h.c.\right]
\end{align}
can be diagonalized via canonical transformation yielding the energy spectrum $\varepsilon_{\pm}(k)=(\frac{k^2}{2m}\!-\!\mu \!)\pm |\Delta_{12}(k) |$. Assuming that all molecules occupy the lowest energy band $\varepsilon_{-}(k)$, the wavefunction corresponding to such state $\ket{\Psi_{\rm FM}}$ is given by Eq.~(5) of the main text.  We calculate the order parameter self-consistently by enforcing the above constraint. Using the variational wavefunction  $\ket{\Psi_{\rm FM}}$, we then calculate the total energy of the system. The contribution of the intralayer interaction reads
\begin{align}
\frac{E^{(\lambda \lambda)}_{\rm int}}{\mathcal{A}}&= \frac 1 8 \sum_{k,q} \left[ V_{\rm eff}^{(\lambda \lambda)} (0)-V_{\rm eff}^{(\lambda \lambda)} (q) \right]  n_{k \lambda } n_{k\!+\!q \lambda} \\
&=\frac{ 1}{8}\! \int \! \frac{d^2\bm q}{(2\pi)^2} \left[ V_{\rm eff}^{(\lambda \lambda)} (0)-V_{\rm eff}^{(\lambda \lambda)} (q) \right]  I(q, \sqrt 2 k_F),\nonumber
\end{align}
and the total energy in the interlayer-coherent superfluid phase is given by
\begin{align}\label{eq:FM}
\frac{E_{\rm FM}}{\mathcal{A}}\! &=\! \frac{4\pi n_0^2}{m} \left[1 - r_s F(\sqrt{8}k_F l_z) \right] \\
&+\frac{16}{\pi}n_0^2\! \int_0^1\! \! x dx \left[ V_{\rm eff}^{(11)} (0)\!-\!V_{\rm eff}^{(11)} (\sqrt 8 k_F \rho_{\Delta} x) \right] \! f(x)\nonumber\\
&= \frac{4\pi n_0^2}{m} \left[1 - r_s F(\sqrt{8}k_F l_z)+\frac{r_s}{2}\frac{Z(\sqrt 8 k_F \rho_{\Delta})}{k_F \rho_{\Delta} \frac{d_{\rm eff}^2}{d_t^2}} \right]. \nonumber
\end{align}
Here the function $F(a)$ is defined as
\begin{align}
F(a)&=32 \sqrt{2\pi} \!\! \int_0^1\! x^2 \! dx \! \left(\arccos(x)\!-\!x\sqrt{1\!-\!x^2}\right)\!e^{-a x}\nonumber\\
&=\frac{32\sqrt{2\pi}}{3a^4}\left[a (3 \pi \!-\! 4 a\!+\! 15 \pi I_0(a)) \!-\!
  3 \pi (12 \!+\! a^2) I_1(a) \right. \nonumber\\
  & \left. + 3 \pi (2 + a^2) L_1(a) - 15 \pi a L_2(a)\right]
\end{align}
with $I_n(x)$ and $L_n(x)$ being the modified Bessel and Struve functions, respectively. As follows from Eq.\eqref{eq:FM}, the magnitude of the intralayer and interlayer interactions depend on different physical parameters $k_F l_z$ and $k_F \rho_{\Delta}$. Thus, one can control their relative contributions to the total energy so that the isospin ferromagnetic state is favorable.

%\section{Variational energy in the state $M_z\neq 0$.}

We conclude this section by calculating the energy of the system in the Ising ferromagnetic state described by the wavefunction $\Psi_{\rm IM}=\prod_{k<\sqrt{2}k_F} c^\dag_{1k}\ket{0}$. Such state has the pseudospin magnetization $\bm M$ aligned along the $z$-axis, which corresponds to all molecules being in a single layer. The intralayer interaction energy in this state can be written as
\begin{align}
\frac{E^{\rm (IM)}_{\rm int}}{\mathcal{A}}&= \frac 1 2 \sum_{k,q}\left[V_{\rm eff}^{(11)} (0)-V_{\rm eff}^{(11)} (q)\right] n_{k 1} n_{k\!+\!q 1} \\
&=\frac{ 1}{2}\! \int \! \frac{d^2\bm q}{(2\pi)^2}\! \left[ V_{\rm eff}^{(\lambda \lambda)} (0)-V_{\rm eff}^{(\lambda \lambda)} (q) \right] \! I(q, \sqrt 2 k_F)\nonumber\\
&= \frac{32}{\pi}n_0^2\! \int_0^1 \! x dx\! \left[ V_{\rm eff}^{(11)} (0)\!-\!V_{\rm eff}^{(11)} (\sqrt 8 k_F\rho_{\Delta} x)\! \right]\!\! f(x). \nonumber
\end{align}
The total energy of the system in the Ising ferromagnetic state reads
\begin{align}
\frac{E^{\rm (IM)}_{\rm t}}{\mathcal{A}}\!&\!=\!\frac{4\pi n_0^2}{m} \!+\!  \frac{32}{\pi}n_0^2\! \int_0^1 \! x dx\! \left[ V_{\rm eff}^{(11)} (0)\!-\!V_{\rm eff}^{(11)} (\sqrt 8 k_F\rho_{\Delta} x) \right]\!\! f(x)\nonumber\\
&\!=\!\frac{4\pi n_0^2}{m}\left(1+r_s\frac{Z(\sqrt 8 k_F \rho_{\Delta})}{k_F \rho_{\Delta} \frac{d_{\rm eff}^2}{d_t^2}}\right).
\end{align}
For typical parameters considered here, the Ising pseudospin ferromagnetic state $\ket{\Psi_{\rm IM}}$  has higher energy than the state $\ket{\Psi_{\rm FM}}$, and thus, can be ignored.

%======================================================

\bibliographystyle{naturemag}
%\bibliography{refs_polar}

\begin{thebibliography}{23}
\expandafter\ifx\csname natexlab\endcsname\relax\def\natexlab#1{#1}\fi
\expandafter\ifx\csname bibnamefont\endcsname\relax
  \def\bibnamefont#1{#1}\fi
\expandafter\ifx\csname bibfnamefont\endcsname\relax
  \def\bibfnamefont#1{#1}\fi
\expandafter\ifx\csname citenamefont\endcsname\relax
  \def\citenamefont#1{#1}\fi
\expandafter\ifx\csname url\endcsname\relax
  \def\url#1{\texttt{#1}}\fi
\expandafter\ifx\csname urlprefix\endcsname\relax\def\urlprefix{URL }\fi
\providecommand{\bibinfo}[2]{#2}
\providecommand{\eprint}[2][]{\url{#2}}

\bibitem[{\citenamefont{Greiner et~al.}({2003})\citenamefont{Greiner, Regal,
  and Jin}}]{Greiner_nature'03}
\bibinfo{author}{\bibfnamefont{M.}~\bibnamefont{Greiner}},
  \bibinfo{author}{\bibfnamefont{C.}~\bibnamefont{Regal}}, \bibnamefont{and}
  \bibinfo{author}{\bibfnamefont{D.}~\bibnamefont{Jin}},
  \bibinfo{journal}{{Nature}} \textbf{\bibinfo{volume}{{426}}},
  \bibinfo{pages}{{537}} (\bibinfo{year}{{2003}});
%
%\bibitem[{\citenamefont{Chin et~al.}({2004})\citenamefont{Chin, Bartenstein,
%  Altmeyer, Riedl, Jochim, Denschlag, and Grimm}}]{Chin_science'04}
  \bibinfo{author}{\bibfnamefont{C.}~\bibnamefont{Chin {\it et al.}}},
%  \bibinfo{author}{\bibfnamefont{M.}~\bibnamefont{Bartenstein}},
%  \bibinfo{author}{\bibfnamefont{A.}~\bibnamefont{Altmeyer}},
%  \bibinfo{author}{\bibfnamefont{S.}~\bibnamefont{Riedl}},
%  \bibinfo{author}{\bibfnamefont{S.}~\bibnamefont{Jochim}},
%  \bibinfo{author}{\bibfnamefont{J.}~\bibnamefont{Denschlag}},
%  \bibnamefont{and} \bibinfo{author}{\bibfnamefont{R.}~\bibnamefont{Grimm}},
  \bibinfo{journal}{{Science}} \textbf{\bibinfo{volume}{{305}}},
  \bibinfo{pages}{{1128}} (\bibinfo{year}{{2004}});
%
%\bibitem[{\citenamefont{Zwierlein et~al.}({2005})\citenamefont{Zwierlein,
%  Abo-Shaeer, Schirotzek, Schunck, and Ketterle}}]{Zwierlein_nature'05}
  \bibinfo{author}{\bibfnamefont{M.}~\bibnamefont{Zwierlein {\it et al.}}},
%  \bibinfo{author}{\bibfnamefont{J.}~\bibnamefont{Abo-Shaeer}},
%  \bibinfo{author}{\bibfnamefont{A.}~\bibnamefont{Schirotzek}},
%  \bibinfo{author}{\bibfnamefont{C.}~\bibnamefont{Schunck}}, \bibnamefont{and}
%  \bibinfo{author}{\bibfnamefont{W.}~\bibnamefont{Ketterle}},
  \bibinfo{journal}{{Nature}} \textbf{\bibinfo{volume}{{435}}},
  \bibinfo{pages}{{1047}} (\bibinfo{year}{{2005}}).

\bibitem[{\citenamefont{Greiner et~al.}(2002)\citenamefont{Greiner, Mandel,
  Esslinger, Hansch, and Bloch}}]{Greiner_nature'02}
  \bibinfo{author}{\bibfnamefont{M.}~\bibnamefont{Greiner {\it et al.}}},
%  \bibinfo{author}{\bibfnamefont{O.}~\bibnamefont{Mandel}},
%  \bibinfo{author}{\bibfnamefont{T.}~\bibnamefont{Esslinger}},
%  \bibinfo{author}{\bibfnamefont{T.}~\bibnamefont{Hansch}}, \bibnamefont{and}
%  \bibinfo{author}{\bibfnamefont{I.}~\bibnamefont{Bloch}},
  \bibinfo{journal}{Nature} \textbf{\bibinfo{volume}{415}}, \bibinfo{pages}{39}
  (\bibinfo{year}{2002}).

\bibitem[{\citenamefont{Ni et~al.}(2008)\citenamefont{Ni, Ospelkaus,
  de~Miranda, Pe'er, Neyenhuis, Zirbel, Kotochigova, Julienne, Jin, and
  Ye}}]{science_polar_KRB'08}
\bibinfo{author}{\bibfnamefont{K.~K.} \bibnamefont{Ni {\it et al.}}},
%  \bibinfo{author}{\bibfnamefont{S.}~\bibnamefont{Ospelkaus}},
%  \bibinfo{author}{\bibfnamefont{M.~H.~G.} \bibnamefont{de~Miranda}},
%  \bibinfo{author}{\bibfnamefont{A.}~\bibnamefont{Pe'er}},
%  \bibinfo{author}{\bibfnamefont{B.}~\bibnamefont{Neyenhuis}},
%  \bibinfo{author}{\bibfnamefont{J.~J.} \bibnamefont{Zirbel}},
%  \bibinfo{author}{\bibfnamefont{S.}~\bibnamefont{Kotochigova}},
%  \bibinfo{author}{\bibfnamefont{P.~S.} \bibnamefont{Julienne}},
%  \bibinfo{author}{\bibfnamefont{D.~S.} \bibnamefont{Jin}}, \bibnamefont{and}
%  \bibinfo{author}{\bibfnamefont{J.}~\bibnamefont{Ye}},
  \bibinfo{journal}{Science} \textbf{\bibinfo{volume}{322}},
  \bibinfo{pages}{231} (\bibinfo{year}{2008});
%
%\bibitem[{\citenamefont{Deiglmayr et~al.}(2008)\citenamefont{Deiglmayr,
%  Grochola, Repp, Mortlbauer, Gluck, Lange, Dulieu, Wester, and
%  Weidemuller}}]{PRL_polar_LiCs'08}
  \bibinfo{author}{\bibfnamefont{J.}~\bibnamefont{Deiglmayr {\it et al.}}},
%  \bibinfo{author}{\bibfnamefont{A.}~\bibnamefont{Grochola}},
%  \bibinfo{author}{\bibfnamefont{M.}~\bibnamefont{Repp}},
%  \bibinfo{author}{\bibfnamefont{K.}~\bibnamefont{Mortlbauer}},
%  \bibinfo{author}{\bibfnamefont{C.}~\bibnamefont{Gluck}},
%  \bibinfo{author}{\bibfnamefont{J.}~\bibnamefont{Lange}},
%  \bibinfo{author}{\bibfnamefont{O.}~\bibnamefont{Dulieu}},
%  \bibinfo{author}{\bibfnamefont{R.}~\bibnamefont{Wester}}, \bibnamefont{and}
%  \bibinfo{author}{\bibfnamefont{M.}~\bibnamefont{Weidemuller}},
  \bibinfo{journal}{Phys.\ Rev.\ Lett.} \textbf{\bibinfo{volume}{101}},
  \bibinfo{pages}{133004} (\bibinfo{year}{2008});
%
%\bibitem[{\citenamefont{Lang et~al.}(2008)\citenamefont{Lang, Winkler, Strauss,
%  Grimm, and Denschlag}}]{PRL_polar_Rb2'08}
  \bibinfo{author}{\bibfnamefont{F.}~\bibnamefont{Lang {\it et al.}}},
%  \bibinfo{author}{\bibfnamefont{K.}~\bibnamefont{Winkler}},
%  \bibinfo{author}{\bibfnamefont{C.}~\bibnamefont{Strauss}},
%  \bibinfo{author}{\bibfnamefont{R.}~\bibnamefont{Grimm}}, \bibnamefont{and}
%  \bibinfo{author}{\bibfnamefont{J.~H.} \bibnamefont{Denschlag}},
  \bibinfo{journal}{Phys.\ Rev.\ Lett.} \textbf{\bibinfo{volume}{101}},
  \bibinfo{pages}{133005} (\bibinfo{year}{2008});
%
%\bibitem[{\citenamefont{Ospelkaus et~al.}(2008)\citenamefont{Ospelkaus, Pe/'er,
%  Ni, Zirbel, Neyenhuis, Kotochigova, Julienne, Ye, and
%  Jin}}]{Ospelkaus_natphys'08}
  \bibinfo{author}{\bibfnamefont{S.}~\bibnamefont{Ospelkaus {\it et al.}}},
%  \bibinfo{author}{\bibfnamefont{A.}~\bibnamefont{Pe/'er}},
%  \bibinfo{author}{\bibfnamefont{K.~K.} \bibnamefont{Ni}},
%  \bibinfo{author}{\bibfnamefont{J.~J.} \bibnamefont{Zirbel}},
%  \bibinfo{author}{\bibfnamefont{B.}~\bibnamefont{Neyenhuis}},
%  \bibinfo{author}{\bibfnamefont{S.}~\bibnamefont{Kotochigova}},
%  \bibinfo{author}{\bibfnamefont{P.~S.} \bibnamefont{Julienne}},
%  \bibinfo{author}{\bibfnamefont{J.}~\bibnamefont{Ye}}, \bibnamefont{and}
%  \bibinfo{author}{\bibfnamefont{D.~S.} \bibnamefont{Jin}},
  \bibinfo{journal}{Nat. Phys.} \textbf{\bibinfo{volume}{4}},
  \bibinfo{pages}{622} (\bibinfo{year}{2008}).

\bibitem[{\citenamefont{Micheli et~al.}(2006)\citenamefont{Micheli, Brennen,
  and Zoller}}]{Micheli_natphys'06}
  \bibinfo{author}{\bibfnamefont{A.}~\bibnamefont{Micheli}},
  \bibinfo{author}{\bibfnamefont{G.~K.} \bibnamefont{Brennen}},
  \bibnamefont{and} \bibinfo{author}{\bibfnamefont{P.}~\bibnamefont{Zoller}},
  \bibinfo{journal}{Nat. Phys.} \textbf{\bibinfo{volume}{5}},
  \bibinfo{pages}{341} (\bibinfo{year}{2006});
%
%\bibitem[{\citenamefont{Wang et~al.}(2006)\citenamefont{Wang, Lukin, and
%  Demler}}]{wang_prl'06}
  \bibinfo{author}{\bibfnamefont{D.-W.} \bibnamefont{Wang}},
  \bibinfo{author}{\bibfnamefont{M.~D.} \bibnamefont{Lukin}}, \bibnamefont{and}
  \bibinfo{author}{\bibfnamefont{E.}~\bibnamefont{Demler}},
  \bibinfo{journal}{Phys. Rev. Lett.} \textbf{\bibinfo{volume}{97}},
  \bibinfo{eid}{180413} (\bibinfo{year}{2006});
%
%\bibitem[{\citenamefont{Wang}(2007)}]{wang_prl'07}
  \bibinfo{author}{\bibfnamefont{D.-W.} \bibnamefont{Wang}},
  \bibinfo{journal}{Phys. Rev. Lett.} \textbf{\bibinfo{volume}{98}},
  \bibinfo{pages}{060403} (\bibinfo{year}{2007});
%
%\bibitem[{\citenamefont{B\"{u}chler et~al.}(2007)\citenamefont{B\"{u}chler,
%  Demler, Lukin, Micheli, Prokof'ev, Pupillo, and Zoller}}]{buchler_prl'07}
  \bibinfo{author}{\bibfnamefont{H.~P.} \bibnamefont{B\"{u}chler {\it et al.}}},
%  \bibinfo{author}{\bibfnamefont{E.}~\bibnamefont{Demler}},
%  \bibinfo{author}{\bibfnamefont{M.}~\bibnamefont{Lukin}},
%  \bibinfo{author}{\bibfnamefont{A.}~\bibnamefont{Micheli}},
%  \bibinfo{author}{\bibfnamefont{N.}~\bibnamefont{Prokof'ev}},
%  \bibinfo{author}{\bibfnamefont{G.}~\bibnamefont{Pupillo}}, \bibnamefont{and}
%  \bibinfo{author}{\bibfnamefont{P.}~\bibnamefont{Zoller}},
  \bibinfo{journal}{Phys. Rev. Lett.} \textbf{\bibinfo{volume}{98}},
  \bibinfo{eid}{060404} (\bibinfo{year}{2007});
%
%\bibitem[{\citenamefont{Chang et~al.}(2009)\citenamefont{Chang, Shen, Lai,
%  Chen, and Wang}}]{Chang_pra'09}
  \bibinfo{author}{\bibfnamefont{C.-M.} \bibnamefont{Chang {\it et al.}}},
%  \bibinfo{author}{\bibfnamefont{W.-C.} \bibnamefont{Shen}},
%  \bibinfo{author}{\bibfnamefont{C.-Y.} \bibnamefont{Lai}},
%  \bibinfo{author}{\bibfnamefont{P.}~\bibnamefont{Chen}}, \bibnamefont{and}
%  \bibinfo{author}{\bibfnamefont{D.-W.} \bibnamefont{Wang}},
  \bibinfo{journal}{Phys. Rev. A} \textbf{\bibinfo{volume}{79}},
  \bibinfo{pages}{053630} (\bibinfo{year}{2009}).

\bibitem[{\citenamefont{Cooper and Shlyapnikov}(2009)}]{cooper_prl'09}
  \bibinfo{author}{\bibfnamefont{N.~R.} \bibnamefont{Cooper}} \bibnamefont{and}
  \bibinfo{author}{\bibfnamefont{G.~V.} \bibnamefont{Shlyapnikov}},
  \bibinfo{journal}{Phys.\ Rev.\ Lett.} \textbf{\bibinfo{volume}{103}},
  \bibinfo{pages}{155302} (\bibinfo{year}{2009}).

\bibitem[{\citenamefont{Micheli et~al.}(2007)\citenamefont{Micheli, Pupillo,
  Buchler, and Zoller}}]{Micheli_prb'07}
  \bibinfo{author}{\bibfnamefont{A.}~\bibnamefont{Micheli {\it et al.}}},
%  \bibinfo{author}{\bibfnamefont{G.}~\bibnamefont{Pupillo}},
%  \bibinfo{author}{\bibfnamefont{H.~P.} \bibnamefont{Buchler}},
%  \bibnamefont{and} \bibinfo{author}{\bibfnamefont{P.}~\bibnamefont{Zoller}},
  \bibinfo{journal}{Phys.\ Rev.\ A} \textbf{\bibinfo{volume}{76}},
  \bibinfo{pages}{043604} (\bibinfo{year}{2007});
%
%\bibitem[{\citenamefont{Gorshkov et~al.}(2008)\citenamefont{Gorshkov, Rabl,
%  Pupillo, Micheli, Zoller, Lukin, and Buchler}}]{gorshkov_prl'08}
  \bibinfo{author}{\bibfnamefont{A.~V.} \bibnamefont{Gorshkov {\it et al.}}},
%  \bibinfo{author}{\bibfnamefont{P.}~\bibnamefont{Rabl}},
%  \bibinfo{author}{\bibfnamefont{G.}~\bibnamefont{Pupillo}},
%  \bibinfo{author}{\bibfnamefont{A.}~\bibnamefont{Micheli}},
%  \bibinfo{author}{\bibfnamefont{P.}~\bibnamefont{Zoller}},
%  \bibinfo{author}{\bibfnamefont{M.~D.} \bibnamefont{Lukin}}, \bibnamefont{and}
%  \bibinfo{author}{\bibfnamefont{H.~P.} \bibnamefont{Buchler}},
  \bibinfo{journal}{Phys.\ Rev.\ Lett.} \textbf{\bibinfo{volume}{101}},
  \bibinfo{pages}{073201} (\bibinfo{year}{2008}).

\bibitem[{\citenamefont{Eisenstein and MacDonald}(2004)}]{Eisenstein_Nature'04}
\bibinfo{author}{\bibfnamefont{J.}~\bibnamefont{Eisenstein}} \bibnamefont{and}
  \bibinfo{author}{\bibfnamefont{A.}~\bibnamefont{MacDonald}},
  \bibinfo{journal}{Nature} \textbf{\bibinfo{volume}{432}},
  \bibinfo{pages}{691} (\bibinfo{year}{2004}).

\bibitem[{\citenamefont{Spielman et~al.}(2000)\citenamefont{Spielman,
  Eisenstein, Pfeiffer, and West}}]{speilman_prl'00}
  \bibinfo{author}{\bibfnamefont{I.~B.} \bibnamefont{Spielman {\it et al.}}},
%  \bibinfo{author}{\bibfnamefont{J.~P.} \bibnamefont{Eisenstein}},
%  \bibinfo{author}{\bibfnamefont{L.~N.} \bibnamefont{Pfeiffer}},
%  \bibnamefont{and} \bibinfo{author}{\bibfnamefont{K.~W.} \bibnamefont{West}},
  \bibinfo{journal}{Phys. Rev. Lett.} \textbf{\bibinfo{volume}{84}},
  \bibinfo{pages}{5808} (\bibinfo{year}{2000}).

\bibitem[{\citenamefont{Hadzibabic et~al.}(2006)\citenamefont{Hadzibabic,
  Kruger, Cheneau, Battelier, and Dalibard}}]{Hadzibabic_nature'06}
  \bibinfo{author}{\bibfnamefont{Z.}~\bibnamefont{Hadzibabic {\it et al.} }},
%  \bibinfo{author}{\bibfnamefont{P.}~\bibnamefont{Kruger}},
%  \bibinfo{author}{\bibfnamefont{M.}~\bibnamefont{Cheneau}},
%  \bibinfo{author}{\bibfnamefont{B.}~\bibnamefont{Battelier}},
%  \bibnamefont{and} \bibinfo{author}{\bibfnamefont{J.}~\bibnamefont{Dalibard}},
  \bibinfo{journal}{Nature} \textbf{\bibinfo{volume}{441}},
  \bibinfo{pages}{1118} (\bibinfo{year}{2006}).

\bibitem[{\citenamefont{Napolitano et~al.}(1997)\citenamefont{Napolitano,
  Weiner, and Julienne}}]{Napolitano_pra'97}
  \bibinfo{author}{\bibfnamefont{R.}~\bibnamefont{Napolitano}},
  \bibinfo{author}{\bibfnamefont{J.}~\bibnamefont{Weiner}}, \bibnamefont{and}
  \bibinfo{author}{\bibfnamefont{P.~S.} \bibnamefont{Julienne}},
  \bibinfo{journal}{Phys. Rev. A} \textbf{\bibinfo{volume}{55}},
  \bibinfo{pages}{1191} (\bibinfo{year}{1997}).

\bibitem[{\citenamefont{Moon et~al.}(1995)\citenamefont{Moon, Mori, Yang,
  Girvin, MacDonald, Zheng, Yoshioka, and Zhang}}]{moon_prb'95}
  \bibinfo{author}{\bibfnamefont{K.}~\bibnamefont{Moon {\it et al.}}},
%  \bibinfo{author}{\bibfnamefont{H.}~\bibnamefont{Mori}},
%  \bibinfo{author}{\bibfnamefont{K.}~\bibnamefont{Yang}},
%  \bibinfo{author}{\bibfnamefont{S.~M.} \bibnamefont{Girvin}},
%  \bibinfo{author}{\bibfnamefont{A.~H.} \bibnamefont{MacDonald}},
%  \bibinfo{author}{\bibfnamefont{L.}~\bibnamefont{Zheng}},
%  \bibinfo{author}{\bibfnamefont{D.}~\bibnamefont{Yoshioka}}, \bibnamefont{and}
%  \bibinfo{author}{\bibfnamefont{S.-C.} \bibnamefont{Zhang}},
  \bibinfo{journal}{Phys. Rev. B} \textbf{\bibinfo{volume}{51}},
  \bibinfo{pages}{5138} (\bibinfo{year}{1995}).

\bibitem[{\citenamefont{Zheng et~al.}(1997)\citenamefont{Zheng, Ortalano, and
  Das~Sarma}}]{zheng_prl'97}
  \bibinfo{author}{\bibfnamefont{L.}~\bibnamefont{Zheng}},
  \bibinfo{author}{\bibfnamefont{M.~W.} \bibnamefont{Ortalano}},
  \bibnamefont{and}
  \bibinfo{author}{\bibfnamefont{S.}~\bibnamefont{Das~Sarma}},
  \bibinfo{journal}{Phys. Rev. B} \textbf{\bibinfo{volume}{55}},
  \bibinfo{pages}{4506} (\bibinfo{year}{1997}).

\bibitem[{\citenamefont{Ye}(2009)}]{ye_private_2009}
\bibinfo{author}{\bibfnamefont{J.}~\bibnamefont{Ye}} (\bibinfo{year}{2009}),
  \bibinfo{note}{private communication}.

\end{thebibliography}

\begin{thebibliography}{10}

\bibitem{Micheli_prb'07} A. Micheli et al., Phys. Rev. A 76, 043604 (2007);

\bibitem{gorshkov_prl'08}
A. V. Gorshkov et al., Phys. Rev. Lett. 101, 073201 (2008).

\bibitem{chan_arxiv'09}
C.-K. Chan, C. Wu, W.-C. Lee, and S. Das Sarma,  Phys. Rev. A 81, 023602 (2010)

\bibitem{cooper_prl'09}
N. R. Cooper and G. V. Shlyapnikov, Phys. Rev. Lett. 103, 155302 (2009).



\end{thebibliography}

\end{document}